\documentclass[aps,prx,reprint,groupedaddress,amsmath,amssymb]{revtex4-2}
\usepackage{xcolor}
\usepackage{graphicx}
\usepackage{braket}
\usepackage{float}
\usepackage{units} 
\usepackage{soul} 
\usepackage{verbatim} 	
\usepackage[hidelinks]{hyperref}

\begin{document}

\title{Reinforcement Learning in Ultracold Atom Experiments}

\author{Malte Reinschmidt}
\author{J\'{o}zsef Fort\'{a}gh}
\author{Andreas G\"unther}
\email{a.guenther@uni-tuebingen.de}
\affiliation{Center for Quantum Science, Physikalisches Institut, 
Eberhard Karls Universit\"at T\"ubingen, Auf der Morgenstelle 14, 
D-72076 T\"ubingen, Germany}
\author{Valentin Volchkov}
\email{valentin.volchkov@tuebingen.mpg.de}
\affiliation{Max Planck Institute for Intelligent Systems, Max-Planck-Ring 4, D-72072 T\"ubingen, Germany}

\begin{abstract}
\noindent
Cold atom traps are at the heart of many quantum applications in science and technology. 
The preparation and control of atomic clouds involves complex optimization processes, that could be supported and accelerated by machine learning. 
In this work, we introduce reinforcement learning to cold atom experiments and demonstrate a flexible and adaptive approach to control a magneto-optical trap. 
Instead of following a set of predetermined rules to accomplish a specific task, the objectives are defined by a reward function.
This approach not only optimizes the cooling of atoms just as an experimentalist would do, but also enables new operational modes such as the preparation of pre-defined numbers of atoms in a cloud. 
The machine control is trained to be robust against external perturbations and able to react to situations not seen during the training. 
Finally, we show that the time consuming training can be performed in-silico using a generic simulation and demonstrate successful transfer to the real world experiment.
\end{abstract} 

\date{\today}

\maketitle

\noindent
Laser cooling and magneto-optical trapping of neutral atoms has been one of the major breakthroughs in science in the last decades \cite{Phillips1998}. 
This enabled access to ultracold atoms and quantum matter with many applications in quantum simulation, computation and sensing \cite{Acin18}.  
Narrow-line cooling lies at the heart of a new generation of magneto-optical traps (MOT) for non-alkali elements and molecules \cite{Schreck2021}, leading to novel quantum matter \cite{Kadau2016,Norcia2021} and vastly improving optical clocks \cite{Bothwell2022}. 
Nowadays, the MOT is the cornerstone for all applications and devices based on ultracold atoms. 
Despite its simple design, operation  
can be quite intricate, with the behavior of the atoms being determined by the interplay of laser cooling, optical pumping, and light-induced interactions. 
For example, precise temporal control of laser parameters and magnetic fields are required for implementation of cooling schemes such as compressed MOT \cite{Petrich94}, molasses \cite{Lett88}, or Raman cooling \cite{Kasevich92}.
With the advent of non-alkali and molecular MOTs also the number of lasers used is rising.
This increasing complexity can create difficulties in controlling and optimizing the production of ultracold samples; however, it can also provide major advantages. 
The key challenge is to handle and make use of this complexity. 
One promising approach is the application of machine learning: methods like Bayesian optimization \cite{Bayes_opt16} have been used to improve control sequences of different phases of an experiment \cite{Tranter2018}, even creating a BEC \cite{Wigley2016,AIcoldatoms3,AIcoldatoms1,ML_BEC22}. 
However, these approaches do not allow for direct feedback during a sequence and cannot react to perturbations and drifts of external conditions.

\begin{figure*}[ht!]
    \includegraphics[width=0.44\textwidth]{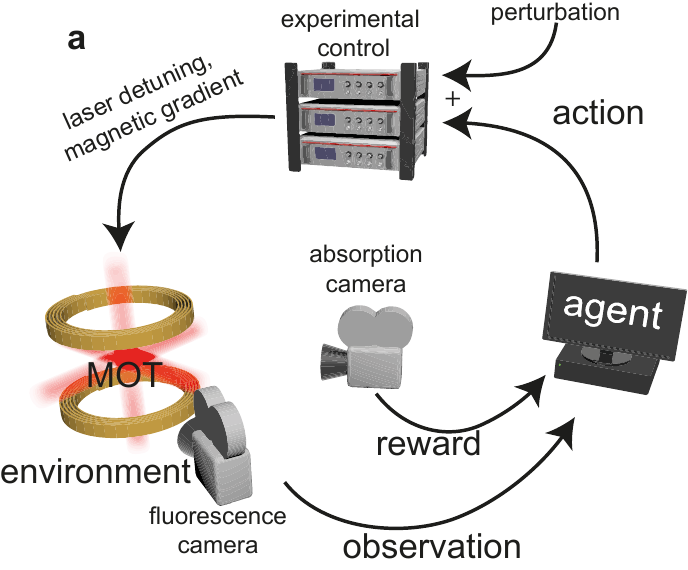} 
    \includegraphics[width=0.55\textwidth]{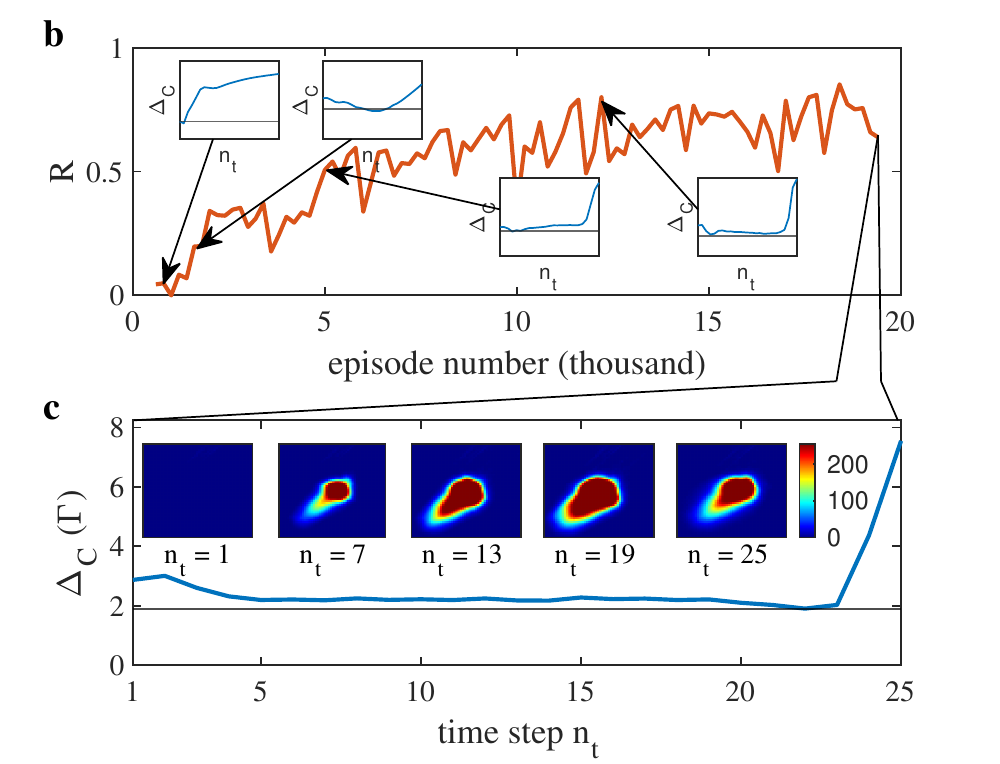}
	\caption{\textbf{Reinforcement Learning for MOT.} 
	\textbf{a} Schematic illustration of the interaction between an RL agent and a MOT-system. 
	The agent takes control (action) via dedicated control parameters, which it freely adjusts during the MOT operation. 
	At each time step $n_t$, the action is based upon the observables (e.g. fluorescence images) given to the agent and the knowledge acquired during the training phase. 
	The latter is based on a reward value $R$, extracted at the end of the MOT cycle (episode) from an absorption image of the atom cloud, which the agent aims to maximize.
    \textbf{b} Evolution of reward $R$ (normalized to reference sequences, as described in the Methods) during a successful training.
    The insets show the control parameter sequences at four points during the training phase.
    Animated fluorescence images corresponding to the episodes in the insets are provided in Supplementary Movie 1.
    \textbf{c} Sequence of control parameter ($\Delta_{\mathrm{C}}$) over the course of an episode for an agent trained to maximize $N/T$. 
    The horizontal line indicates the value for maximal loading rate of the MOT.
    The inset shows the fluorescence images observed by the agent at various time steps $n_t$ during the episode.
  }
\label{fig:1}
\end{figure*}

In the last decade, reinforcement learning (RL) \cite{Sutton1998}  has shown human-level performances in a range of tasks that were previously thought to be impossible for machines to complete. 
While impressive success was achieved in games and simulated environments \cite{Mnih2015,AlphaZero2018,AlphaGoZero2017,AlphaStar2019,Diplomacy22}, application of RL to real world systems remained challenging \cite{Dulac-Arnold2021}. 
In recent years, control of various physical systems was established through RL algorithms, leading to a paradigm shift in how control is approached and implemented \cite{RL_QuantSystem22,Tunnermann:19,Konishi2022,Bellemare2020,Degrave2022}. 
Generally speaking, in RL the control algorithm denoted as \textit{agent} interacts with an environment based on observations. 
As a result of its actions the agent receives rewards, which are used in a training algorithm, allowing the agent over time to learn how to maximize the cumulative reward.
The task is defined directly through the reward function, leaving the specific control sequence solely to the agent. 
So-called deep RL \cite{deepRLduan16} incorporates deep artificial neural networks allowing for model-free control as a function of many, potentially high dimensional inputs, such as images.

Here, we introduce deep RL to ultracold atom experiments by means of controlling a MOT-system.
We enhance an existing cold atom apparatus used to cool rubidium ($^{87}\mathrm{Rb}$) atoms by an RL agent, taking advantage of the fluorescence light, which is a by-product of laser cooling. 
In most similar apparatuses, fluorescence images are only a basic diagnostic tool, whereas they can contain a complex and non-linear representation of the system and are thus an ideal candidate for input to an RL agent. 
We use an off-the-shelf implementation \cite{rlcoach} of a popular RL algorithm  with minimal modifications to the code, running on a desktop computer.
As proof-of-concept, we perform training of the agent directly on the experiment, which rapidly converges to a cooling scheme resembling well known molasses cooling.
Furthermore, we train the agent to be capable to react to sudden changes and observe generalization to situations that are not included during the training.
We demonstrate new modes of operations of the MOT, such as loading predefined number of atoms by engineering the corresponding reward function. 
Finally, we explore how an agent trained on a simple, neural network-based MOT simulator can be transferred to the real world experiment, opening the avenue to fast pre-training, extensive parameter optimization and large number of control parameters.

\section*{Setup}
\subsection*{Experimental System}
The demonstration of RL-based MOT control is implemented on a cold atom apparatus, that employs many standard techniques (see Methods).
Similar setups can be found in many labs around the world \cite{everycoldatom}.
Like all experiments with cold atoms, these experiments are performed in recurring cycles, each starting with the preparation of an atomic cloud in a MOT. 
We dedicate the present manuscript to this preparation phase and conclude the experimental cycle directly thereafter by detecting the collected atoms.
For that, we use time-of-flight absorption imaging  \cite{Ketterle1999}, in which the expanding cold atom cloud is illuminated with a resonant laser beam and the beam transmission measured on a CCD camera. 
This image is used to determine the final state of the atomic cloud, consisting of the number of atoms $N$ and their average kinetic energy, which we denote by 'temperature'  $T$.
In our specific system, after a loading time of \unit[1.5]{s}, the number of atoms typically amounts to $N\approx 1.5\times 10^8$ at a temperature of about \unit[100]{$\mu K$}.

\subsection*{Implementation of the control}
While most of the experimental control is running on a separate system, the agent algorithm is implemented on a desktop PC with access to fluorescence images and control over specific experimental parameters (see Methods).
We limit the parameters controlled by the agent to the frequency detuning $\Delta=\omega_0-\omega_L$ of the cooling laser at frequency $\omega_L$ with respect to the atomic resonance at $\omega_0$ and the magnetic field gradient $B'$ at the MOT center along the field's symmetry axis (see Methods). 
They are among the most influential parameters, both on the number of atoms and the temperature in the MOT, and are typically varied over the course of the MOT loading. 
In this study, the RL agent is allowed to continuously adjust one or both of these parameters. 
The detuning $\Delta$  is restricted to the range of \unit[0]--\unit[8.25]$\Gamma$ with $\Gamma$=2$\pi\times$\unit[6.063]{MHz} being the cooling transition linewidth \cite{Gutterres02}. 
The magnetic field gradient $B'$ can be varied by the agent between \unit[7.5]--\unit[22.5]{G/cm}.

Each MOT cycle of \unit[1.5]{s} duration constitutes an episode and is divided into 25 sequential time steps $n_t$ of \unit[60]{ms} length.
Figure \ref{fig:1}\textbf{a} illustrates the main principle for controlling the MOT with an RL agent.
The MOT is part of the environment, which denotes the entire physical system, including the atoms, the surrounding experimental set-up and the lab.
Within each time step a vector of observables consisting of fluorescence images and additional parameters such as current time step index and values of control parameters is passed as input to the agent.
The agent processes the input using a trainable neural network and determines the next action, i.e. computes the output values for the controlled experimental parameters.
Subsequently, the control parameters are applied through the experimental control and the environment evolves for the duration of the time step.

\subsection*{Training of the agent}
In order to train the agent for a specific task, the reward function needs to be defined accordingly. 
In general, a reward can be generated after each time step, in our implementation a reward is attributed only after the last time step when the state of the atomic cloud has been obtained through absorption imaging. 
The essence of RL is to train the agent's policy to maximize the reward. 
The RL algorithm used here is the deep deterministic policy gradient (DDPG) \cite{ddpg}, which has been developed for high dimensional observations (e.g. images)  and is capable of outputting continuous actions, i.e. control parameters.
The training takes place in the course of a number of episodes, during which the policy is continuously updated according to incoming rewards.
During the training, exploration is introduced via noise added to the actions of the agent.  
A single training consists of up to $10^5$ episodes, depending on the number of control parameters and the specific definition of the reward function. 
Figure~\ref{fig:1}\textbf{b} shows a successful training, in the course of which the obtained reward increases and saturates, while the output values (control parameters) converge to a stable sequence (see insets). 
Figure~\ref{fig:1}\textbf{c} shows a typical control sequence within one episode, together with representative fluorescence images. 

A well known problem of learning algorithms is that some inputs may be ignored and instead "shortcuts" are memorized \cite{Shortcuts}. 
In our scenario, this approach is particularly tempting as the experimental setups are typically very stable and run reproducibly, thus operation with static control sequences is very common. 
To prevent this we introduce a training perturbation, i.e. an offset, that is randomly generated at the start of each episode and added to the agent's output value of the control parameter, effectively shifting the experimental conditions. 
Such a perturbation is equivalent to a long-term drift of the control parameter that may occur in experiments due to changes in the environment (e.g. lab temperature).
As the offset value is unknown to the agent, it is forced to extract information from the fluorescence images in order to compute the control parameter. 
This approach is similar to \textit{domain randomization} that induces variability during training on simulations, making the agent more robust in real world environments \cite{domainRand17}. 

\begin{figure*}[ht!]
    \includegraphics[width=0.95\textwidth]{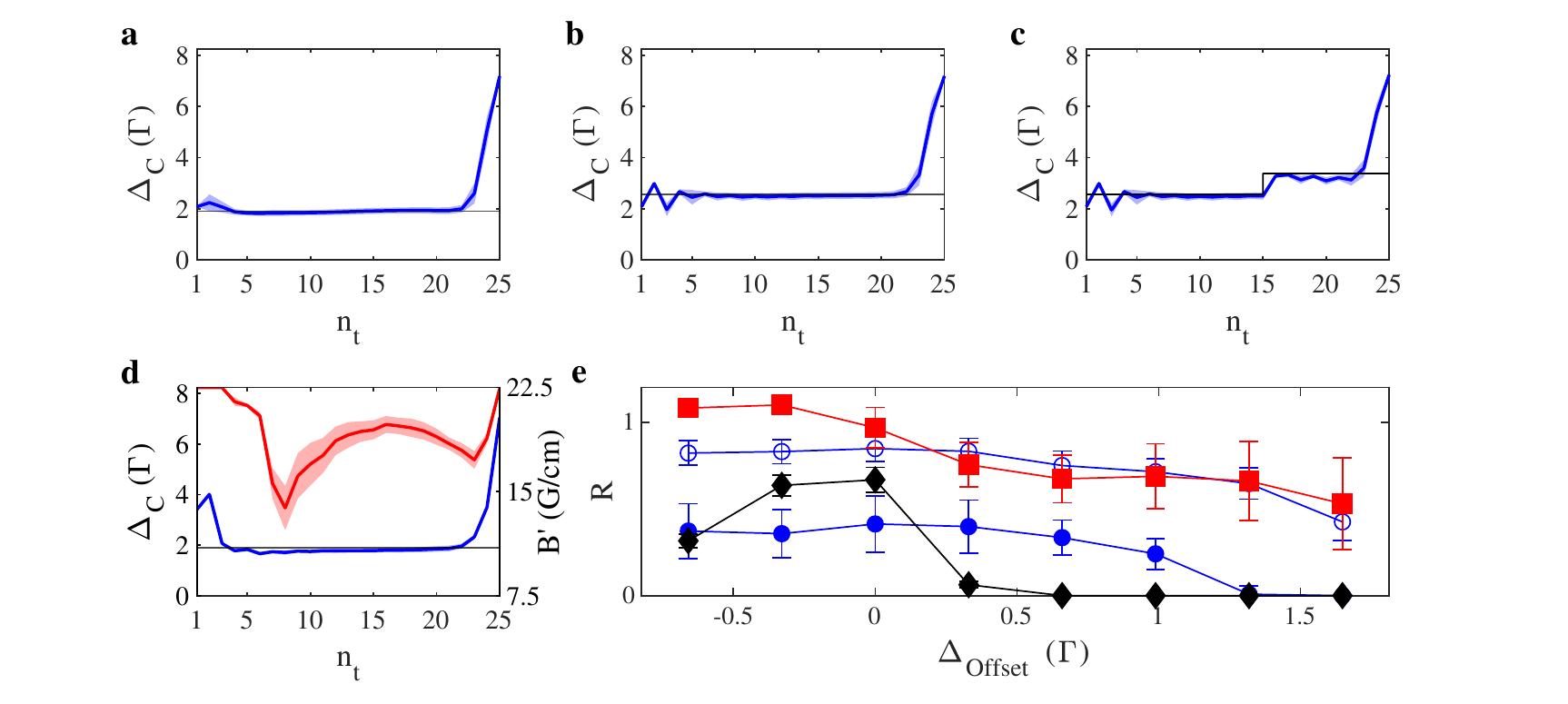}
	\caption{\textbf{Evaluation of agent trained to optimize $N/T$.} 
        \textbf{a} Control parameter sequence over the course of episodes with mean value (colored solid line) and standard deviation (transparent region). 
        Data are obtained by an agent evaluated at $\Delta_{\mathrm{Offset}}=0$. The horizontal black line indicates the point of optimal loading $\Delta_{\text{opt}}$.
        \textbf{b} Same agent as in (a), but evaluated at an offset $\Delta_{\mathrm{Offset}}= \unit[0.66]{\Gamma}$. 
		The optimal loading line has been shifted accordingly.
        \textbf{c} Same agent as in (b), but evaluated with a sudden step in the $\Delta_{\mathrm{Offset}}$ at $n_t=15$ with amplitude \unit[0.82]{$\Gamma$}.  
        \textbf{d} Control parameter sequence over the course of episodes for an agent controlling two output parameters, corresponding to the cooling laser detuning $\Delta_{\mathrm{C}}$ (blue) and the magnetic field gradient $B'$ (red).
        \textbf{e} Reward values (mean and standard deviation) of three different agents as extracted during the evaluation for different $\Delta_{\mathrm{Offset}}$ values:  
        $\textcolor{blue}{\circ}$/$\textcolor{blue}{\bullet{}}$ - agent controlling perturbed detuning ($\Delta_{\mathrm{Offset}}\neq 0$ during training);
        $\blacklozenge{}$ - agent controlling  detuning ($\Delta_{\mathrm{Offset}}=0$ during training);
        $\textcolor{red}{\blacksquare{}}$ - agent controlling $B'$ and perturbed detuning. 
        The $\textcolor{blue}{\bullet{}}$ measurements include a sudden change in the $\Delta_{\mathrm{Offset}}$ value at $n_t=15$ during evaluation, as illustrated in \textbf{c}.
        The timing and amplitude of this "step" in the offset was the same for all detuning offsets.}
\label{fig:2}
\end{figure*}

\subsection*{Evaluating the agents performance}
In order to evaluate and compare the performance of trained agents, we define a testing procedure, aiming for the agent's ability to adapt to different experimental conditions, as well as the reproducibility of the results when presented with the same conditions. 
As such we test the agent by adding different predetermined perturbation offset values (unknown to the agent) to the control parameter at the start of an episode, repeating the test for each of these values up to 200 times. 
We then compare the final results in terms of the number of collected atoms, the achieved cooling, and the corresponding reward. 
Additionally, we study to which policy the agent has converged by analyzing the sequences and spread of the output values over the course of multiple episodes for each of the different offset values \cite{note1}.

\section*{Results}
\subsection*{Standard MOT operation mode}
In many cold atom experiments MOT operation aims for maximizing the number of trapped atoms while reaching the lowest possible temperature, as it is one of the keys towards quantum degeneracy. 
For a proof-of-principle demonstration we define the reward as $R\propto N/T$ and focus first on a single control parameter -- the laser detuning $\Delta$. 
The physical value of the laser detuning $\Delta$ is determined through the agent's control value $\Delta_{\mathrm{C}}$ and the perturbation offset $\Delta_{\mathrm{Offset}}$ (unknown to the agent) via $\Delta = \Delta_{\mathrm{C}}  - \Delta_{\mathrm{Offset}}$. 
The magnetic field gradient is set to $B'=\unit[15]{G/cm}$.

With the MOT principles being based on velocity dependent light forces, the loading rate of atoms strongly depends on laser detuning and shows a maximum at a specific value. 
In our experiment we measured this value to $\Delta_{\mathrm{opt}} =$ \unit[1.9]{$\Gamma$}. 
The temperature is known to approximately scale inversely proportional to $\Delta$ \cite{Dalibard89}. 
The final temperature is then mainly given by the value of $\Delta$ applied at the end of the episode. 
With this in mind, the intuitive optimal policy consists of first going to the point of highest loading rate, before rapidly ramping $\Delta$ to the maximum value at the end of the episode to reduce the temperature. 
This policy corresponds to what is commonly used in labs working with MOTs and provides a reference for studying the application of RL to MOT control (see Methods). 

In Fig.~\ref{fig:1}\textbf{b}, we present the learning progress: the agent discovers early on that going to large detunings at the end of an episode results in a massive boost in the reward, as witnessed in the first insets.
This is consistent with the fact that reward is issued only in the last time step.
During the second half of the training, the agent fine-tunes the loading phase.
The control output sequences $\Delta_{\mathrm{C}}$ for a trained agent evaluated at two different detuning offsets are shown in Fig.~\ref{fig:2}\textbf{a} and \textbf{b}. 
We observe, that the sequence of the detuning $\Delta_{\mathrm{C}}$ over the course of an episode for a trained agent matches our expected intuitive policy. 
While the first fluorescence images of each episode are blank, as no atoms have yet been loaded into the MOT, the agent develops an initial search pattern, jumping to different values of $\Delta_{\mathrm{C}}$. 
Once the first atoms are trapped and visible on the fluorescence image, the agents compensates for the perturbation offset $\Delta_{\mathrm{Offset}}$ and sets the output value corresponding to the highest loading rate, as shown in Fig.~\ref{fig:2}\textbf{b}. 
This allows the agent to load as many atoms into the MOT as possible, before increasing $\Delta_{\mathrm{C}}$ towards the end of the episode in order to reduce the temperature and maximize the reward. 
This control policy is remarkably stable, as seen in the small standard deviations in Fig.~\ref{fig:2}\textbf{a} and \textbf{b}. 
Fig.~\ref{fig:2}\textbf{e} summarizes the performance in terms of the achieved rewards for several trained agents, evaluated at different offsets.
At large offsets, the drop in the obtained reward is due to an increased number of time steps needed to locate the optimal loading point. 

For sake of consistency, we also trained an agent without perturbations on the control values. 
This agent learns the optimal policy, however, it ignores the changes related to $\Delta_{\mathrm{Offset}}$ in the observations during the evaluation phase and plays back the same memorized action sequence.
The reward then simply mirrors the loading rate of the MOT at different offsets from the optimal loading point, as can be seen in Fig.~\ref{fig:2}\textbf{e}.

\subsection*{Generalization to unfamiliar situations}
The capacity to adequately react to new situations, unseen during the training, is one of the key interests in RL since it is a witness of the robustness and usefulness of the trained agent. 
Thus, we tested our agent by exposing it to sudden changes in the environment and observing the real-time adaption of its behavior. 
In particular, the offset applied to the output value $\Delta_{\mathrm{C}}$ during evaluation was suddenly increased at $n_t = 15$ by \unit[0.82]{$\Gamma$}. 
Such a shift can be compared to a sudden change in the environment, which may be caused by singular acoustic or electronic events. 
As shown in the control parameter sequence of Fig.~\ref{fig:2}\textbf{c}, the agent was able to compensate for this shift after just a single step, i.e. based on a single fluorescence image.
This demonstrates the ability of the trained agent to generalize and adapt to conditions not present during the training. 
Note, that the corresponding reward values ($\textcolor{blue}{\bullet{}}$ in Fig.~\ref{fig:2}\textbf{e}) cannot be directly compared to those without shift ($\textcolor{blue}{\circ}$ in Fig.~\ref{fig:2}\textbf{e}), as for a short interval of time, before the agent was given a chance to react, the detuning was being shifted close enough to the resonance where loading not only stops, but atoms were actively removed from the trap. 
As a result the total collected number of atoms was reduced, leading to lower reward values.

\begin{figure*}[ht!]
    \includegraphics[width=0.95\textwidth]{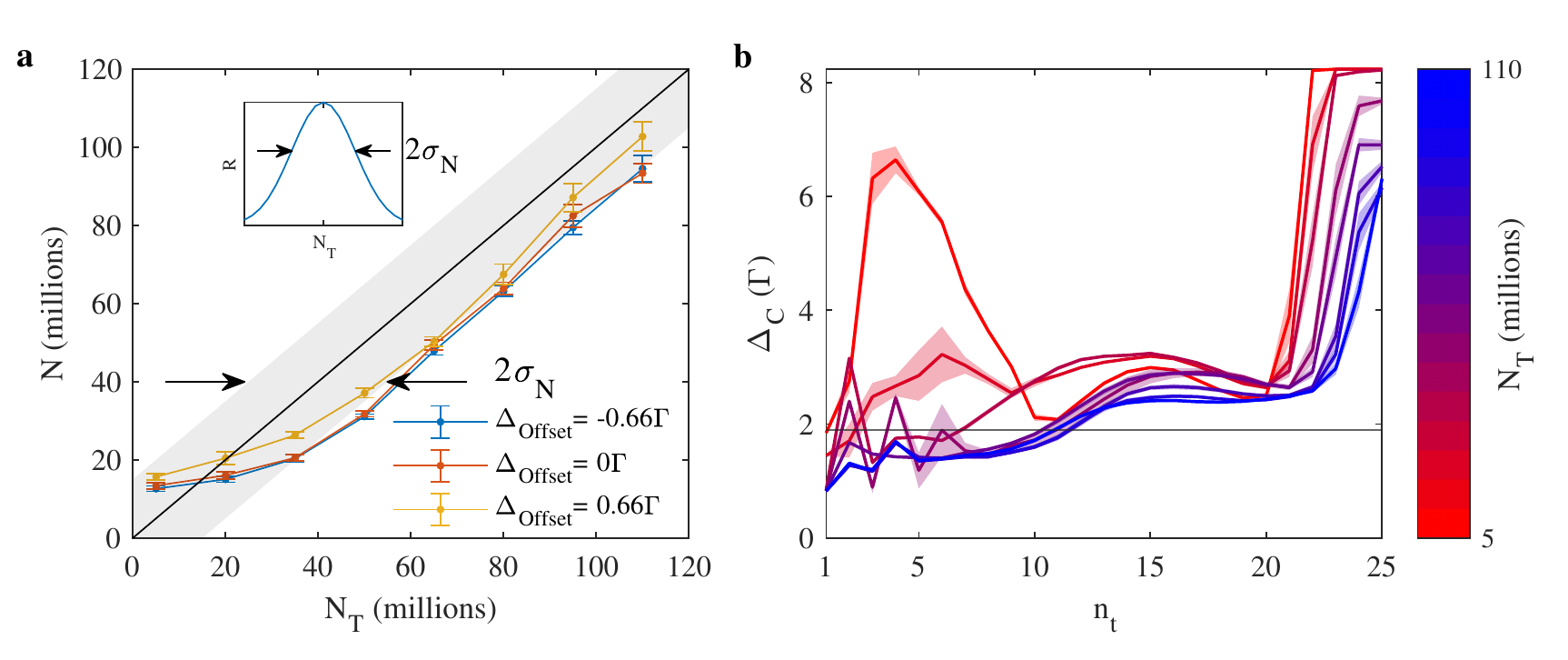}
	\caption{\textbf{Atom number on demand.} 
	\textbf{a} Evaluation for three different offsets in terms of the targeted ($N_{\mathrm{T}}$) and achieved ($N$) number of atoms for an agent trained to load an adjustable amount of atoms. The inset shows the reward function, defined as gaussian profile around $N_{\mathrm{T}}$, with standard deviation $\sigma_N=15\times10^6$. The black line and grey area correspond to the response  $N=N_{\mathrm{T}}\pm\sigma_N$, with mean and standard deviation, respectively.
    \textbf{b} Control parameter sequence for $\Delta_{\text{offset}}=0$ and different vales of $N_{\mathrm{T}}$, ranging from $5\times10^6$ (red) to $110\times10^6$ (blue). Qualitatively similar sequences were observed for non-zero offsets.
  }
\label{fig:3}
\end{figure*}

\subsection*{Increasing action space dimension}
So far, the action space has been limited to a single control parameter. 
In the following we expanded the action space to also include the current that defined the gradient of the magnetic field, i.e.~$B'$, while the definition of the reward as $R\propto N/T$  and other aspects during the training remained unchanged.
The perturbation offset applied to $\Delta$ was found to be sufficient in preventing the learning of shortcuts, so no offset was added to $B'$.

With the controls extended to the gradient of the magnetic field, schemes similar to compressed MOT \cite{Petrich94} or polarization gradient cooling \cite{Dalibard89} could expected to be discovered by the agent. 
However, these techniques require control on time scales much shorter than the time step duration of \unit[60]{ms}. 
As consequence, no intuitive optimal policy is known for this scenario.

We evaluate the trained agent, similar as before, with the resulting control sequences for zero offset shown in Fig.~\ref{fig:2}\textbf{d}. 
The control of the detuning $\Delta$ is very similar to before. 
For $B'$, the agent initially maximizes this control parameter. 
Once the point of highest loading rate is found, $B'$ is lowered. 
In a manner reminiscent of a compressed MOT,  $B'$ is increased to maximum at the end of the episode.
In terms of performance, one could expect an improvement due to the additional control parameter. 
Indeed Fig.~\ref{fig:2}\textbf{e} shows generally better results for this agent.
Interestingly, for $\Delta_{\mathrm{Offset}} = -0.33\Gamma,-0.66\Gamma$ the agent outperformed the reference-sequences in which $B'$ was kept constant. 
This suggests that MOT operation can be further improved, provided the agents is granted control over further experimental parameters.

\begin{figure*}[ht!]
    \includegraphics[width=0.43\textwidth]{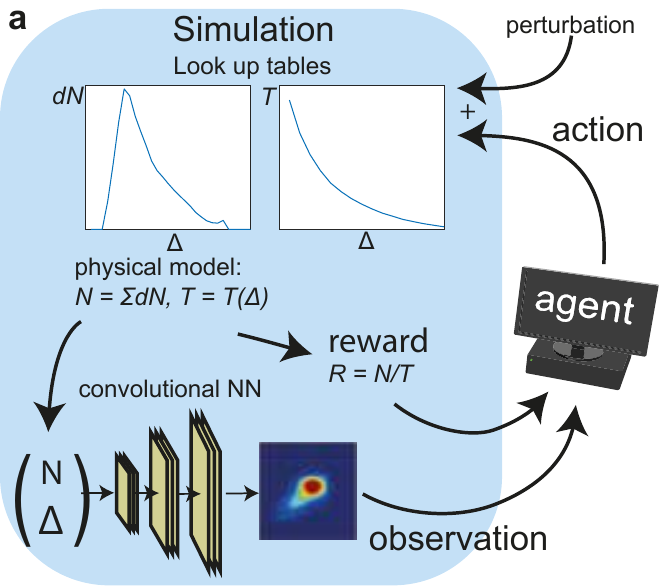}
    \includegraphics[width=0.43\textwidth]{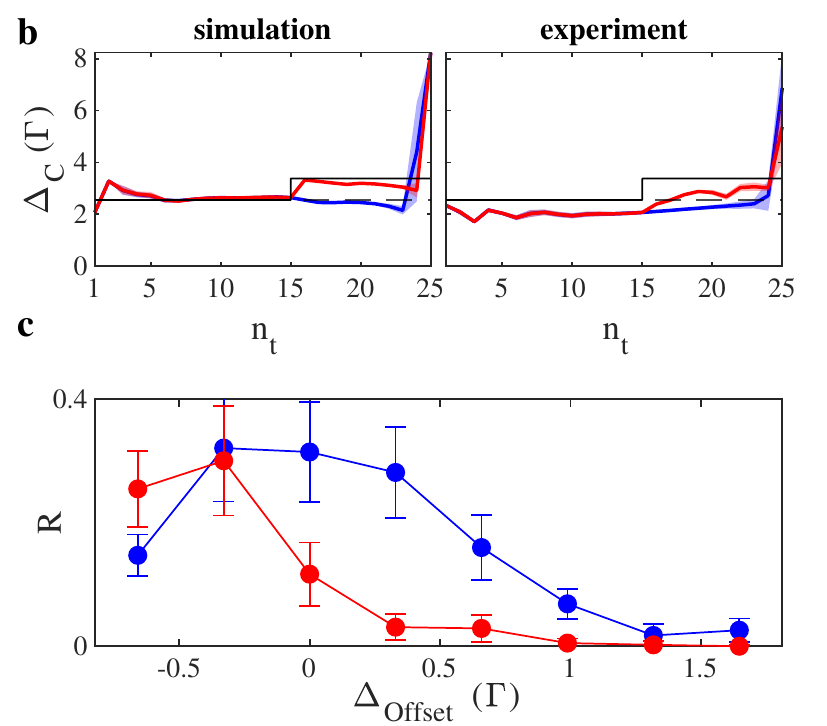}
	\caption{\textbf{Sim-to-real transfer.} 
	\textbf{a} Schematic diagram of the training on the simulation: the graphs illustrate the experimentally measured values for the number of atoms $dN$ loaded during a time step and the temperature $T$ as function of the detuning $\Delta$. At the bottom the generation of fluorescence images is illustrated:  a convolutional neural network takes the number of atoms $N$ and $\Delta$ as input to produce realistically looking images (see Methods).
	\textbf{b} Control parameter sequences of an agent trained on the simulation and evaluated for $\Delta_{\text{offset}}=0.66\Gamma$ on the simulation (left) and experiment (right). The red curves show the resulting behavior if the offset is suddenly changed during the episode (cf. Fig.~\ref{fig:2}\textbf{c}).
	The solid/dashed black line indicate the point of optimal loading with/without the sudden change of the offset.
	\textbf{c} Reward values of agents trained on simulation and evaluated on the experiment. 
	$\textcolor{blue}{\bullet{}}$/$\textcolor{red}{\bullet{}}$ --- reward without/with sudden change of the offset.
  }
\label{fig:4}
\end{figure*}

\subsection*{Reward function engineering}
Now we go beyond the standard task of maximizing $N/T$ and showcase how the engineering of the reward function enables the realization of new modes of operation. 
As an example, one important task is the preparation of a sample with an adjustable number of atoms.

Using RL, we were able to train an agent to achieve variable targeted atom numbers in each episode, while acting only on the detuning of the cooling laser and keeping all other parameters fixed ($B'= \unit[15]{G/cm}$).
For this, we introduce a new parameter $N_{\mathrm{T}}$ for the target number of atoms and pass it to the agent as additional observable. 
The reward function is altered to depend on this parameter, using a Gaussian profile centered around $N_{\mathrm{T}}$ with a width $\sigma_N$. 
To ensure that the atoms remain cold, we include a factor $1/T$ in the reward function:
\begin{equation} 
\label{eq:RNt}
R \propto \frac{1}{T} \exp{\left(-\frac{(N-N_{\mathrm{T}})^2}{2\sigma_N^2}  \right)}
\end{equation}
For the training we limit the range of $N_{\mathrm{T}}$ to  0--120 million atoms and set the acceptance range $\sigma_N$ equal to 15 million atoms. 
$N_{\mathrm{T}}$ is randomly generated for each training episode, with a uniform distribution across the entire range. 
As before, a random perturbation offset was added to the output of the agent at each training episode to prevent the learning of shortcuts. 
The agent must therefore learn to manipulate the loading rate to achieve the requested atom number, while simultaneously compensating the offset.

Intuitively, one possible control strategy could be to divide the episode in two phases: a loading phase with the detuning remaining at the optimal loading point but its length adjusted to meet the requested atom number, followed by a cooling phase at larger detunings.
Alternatively, the loading could take place at variable loading rate but constant length and terminated with a jump to maximal detuning for cooling at the episode's end.

We evaluated the trained agent at given $N_{\mathrm{T}}$ values, ranging from 5--110 million atoms.
When comparing the measured number of atoms to the target values in Fig.~\ref{fig:3}\textbf{a}, we find, that they fall roughly within a region of $\pm\sigma_N$ around $N_{\mathrm{T}}$. 
The agent is thus capable of adapting to different number of atoms in consecutive episodes. 
More importantly, the performance of the agent is robust with respect to the applied perturbation during the evaluation, as demonstrated by the data obtained for non-zero $\Delta_{\mathrm{Offset}}$. 
This robustness is a direct consequence of the real-time feedback of the RL agent and would be impossible to obtain for manual and static control sequences.

The control parameter sequences presented in Fig.~\ref{fig:3}\textbf{b} were applied by the agent for $\Delta_{\mathrm{Offset}}=0$.
They show that the agent varies both: the loading rate (i.e. distance to $\Delta_{\mathrm{opt}}$) as well as the duration of the cooling phase (i.e. jump to larger detuning) depending on $N_{\mathrm{T}}$.
Furthermore, for increasing values of $N_{\mathrm{T}}$ the agent limits the cooling in a trade-off for larger number of atoms.
Indeed, increasing the detuning reduces the confinement of the MOT and leads to an increased loss of atoms.
Due to the absolute nature of the acceptance range $\sigma_N$ in the reward function (\ref{eq:RNt}), the loss of atoms affects the reward only for larger $N_{\mathrm{T}}$.
In principle, the agent could collect a larger number of atoms in order to compensate for this loss and to better meet the target by staying closer to $\Delta_{\mathrm{opt}}$ during the loading phase.
This shows that despite its overall success, the policy of the trained agent is not yet optimal. 

\subsection*{Sim-to-real transfer}
As with other implementations of RL in real world environments, a major challenge is the required training time.
Depending on the complexity of the reward function and the number of parameters, the training performed for this work took \unit[12-48]{h}.
Adding complexity to the reward function, or expanding the system to include more parameters increases this time.

A typical approach to this problem in RL for robotic systems is sim-to-real transfer.
If given a simulation sufficiently close to the environment, general behavior can be studied and hyperparameters of the algorithm being optimized.
Ideally, the algorithm trained on the simulation could then be transferred to the experiment, either directly, or as a basis for an additional shorter training to adapt to the real world.

To investigate sim-to-real transfer, we have implemented a simple phenomenological model of the MOT, as depicted in Fig.~\ref{fig:4}\textbf{a} and detailed in the Methods section.
We modeled the internal state of the MOT assuming that both, the number of atoms $dN$ loaded per time step as well as the temperature $T$ only depend on the current value of $\Delta$ and extracted their functional relation, $dN(\Delta)$ and $T(\Delta)$, from the experiment. 
Furthermore, we trained a generative neural network on experimental data to reproduce fluorescence images corresponding to the simulated MOT state. 
As with the training on the experiment, a perturbation offset was applied to the action in order to force the agent to rely on the fluorescence images. 

Figure~\ref{fig:4}\textbf{b} (left) shows the control sequence of an agent trained and evaluated on the simulated environment, reproducing the behavior of experimentally trained agents.
We then evaluated the simulation-trained agent directly on the real world experiment.
The corresponding control presented in Fig.~\ref{fig:4}\textbf{b} (right) qualitatively mimics the behavior of the agent evaluated on the simulation and thereby indicates a successful sim-to-real transfer.
More importantly, when presented with a sudden shift of the offset value during the episode, the sim-to-real agent adapted to the change in the same manner as the agent trained on the experiment.
The agent was therefore not only able to generalize from its training on the simulations to the real world experiment, but was even able to take this a step further and generalize to unknown situations that did not occur during the training.

The absolute reward values achieved by the sim-to-real agent, shown in Fig.~\ref{fig:4}\textbf{c}, are inferior to those of the agents trained directly on the experiment.
The main reason for this is that the agent did not fully compensate for the perturbation offset when evaluated on the experiment, as can be seen in Fig.~\ref{fig:4}\textbf{b}.
This shows that there is still a considerable gap between the simulations and the experiments.
Possibly, one contributing factor is the smoothing of the synthetic images, lacking high spatial frequency features of the fluorescence images in the experiment (see Methods).
Other factors may be the simplifications in the simulation, which ignore light-induced interactions in the MOT, multiple photon scattering and changing lab conditions. 
By addressing these effects in the simulation, and improving the details in the synthetic images, the sim-to-real gap may be further closed.

We publish the full code needed to run the simulation and encourage the interested reader to try out the RL on a simulated MOT \cite{RLMOT_code}.
Furthermore, the reward function can be easily modified to test new modes of operation of a MOT on a computer. 
Even the phenomenological model can be retrained on custom training data, allowing experimentalists to safely evaluate the possibilities provided by RL without changes to their experimental setup.

\section*{Discussion and Outlook}
We have successfully applied RL control to ultracold atom experiments and demonstrated its applicability by means of a MOT system. 
For this we have implemented a well studied algorithm (DDPG), running on a desktop computer.
Our pioneering work will enable many labs around the world to adapt this approach to their specific needs and extend their research.
In detail we have shown, that an RL agent for MOT control is able to find meaningful control strategies during a training phase and is capable to react in real-time to unexpected events, not seen during the training.
The real-time feedback to the agent through fluorescence images allows for operation under unstable conditions and is a central distinction with respect to other machine learning-optimized ultracold atom experiments.
Providing additional observations such as real-time values of the laser intensity and lab temperature (that tend to fluctuate and drift over time) has the potential to  further improve the performance of such a system.
In this work we have facilitated the convergence of the training to comprehensible control strategies by limiting the number of parameters and their range. 
In future work, increasing the range, giving control to additional parameters (e.g. laser intensities and polarizations) and reducing the duration of the time steps could lead to discovery of new and possibly counter-intuitive operation modes of a MOT \cite{ML_BEC22}. 
At the same time, allowing the agent to control additional parameters would enable the operation of MOTs for non-alkali species and molecules that require multiple lasers and elaborate control sequences. 
The policies discovered by RL agents would have the potential to shed new light into the physics of the light-matter interaction, especially when working with such complex systems.

In addition, we have demonstrated a paradigm change in ultracold atom physics by using reward engineering to focus on the goal of a specific task, instead of manually crafting control sequences.
As an application of this principle, we have trained an agent that is capable of producing samples with user defined number of atoms.
Conceivably, position dependent reward functions could be used to realize complex spatio-temporal control of ultracold samples, reminiscent to RL control of a plasma in a tokamak fusion generator \cite{Degrave2022}.
The promising results of transferring an agent trained on a simple simulation to the real experiment opens multiple opportunities for further research:
the physical model used for the simulation can be improved by incorporating light-induced losses, with the necessary parameters readily acquired on the experiment.  
Domain randomization in the form of manipulation of synthetic fluorescence images could further improve the sim-to-real transfer efficiency.

With the MOT being the workhorse in the field of ultracold atoms, it presents a new, exciting platform for the development and testing of novel RL algorithms. 
Indeed, RL algorithms such as soft actor-critic (SAC) \cite{SAC} or proximal policy optimization (PPO) \cite{PPO} can be readily implemented and tested in our simulation and brought to the experiment.
To foster progress in this new and exciting field, we made the code for training and evaluating an agent based on a simulated MOT publicly available \cite{RLMOT_code}. 
Finally, our work constitutes the first step to the production of degenerate quantum gases using RL control, provided that suitable real-time observables (e.g. ion signals \cite{Guenther2009} or non destructive imaging methods  \cite{phasecontrast97,faraday_img_12}) can be implemented.

\section*{Methods}

\subsection*{MOT operation and implementation}
MOT operation requires a quadrupole magnetic field and in the most common configuration a pair of counter-propagating laser beams (cooling beams) in all three spatial directions. Together, this makes for a position and velocity dependant light force (radiation pressure), which cools and traps neutral atoms from the gaseous phase at the center of the magnetic quadrupole field where the beams are overlapping. 

Our MOT system is based on $^{87}Rb$ atoms and described in \cite{Fortagh2000}. 
The quadrupole field is generated by a pair of coils in Anti-Helmholtz configuration, providing a field gradient of $B'=\unit[14]{G/cm}$ at the quadrupole center, along the coil symmetry axis for about \unit[1]{A} current in the coils. 
The coil system is placed inside a UHV science chamber at a pressure of $\unit[10^{-11}]{mbar}$ and operated via a voltage controlled current source. 
The cooling beams with waists ($1/e^2$-beam radius) of \unit[12]{mm} and powers of about \unit[15]{mW} are circularly polarized and overlap at the center of the magnetic quadrupole between the coils. 
The MOT is loaded from a beam of precooled Rb atoms emitted from a two-dimensional (2D)-MOT-source. 
The latter consists of field coils and thermal rubidium emitter sources (dispensers), providing a rubidium partial pressure of $>\unit[10^{-9}]{mbar}$ in the corresponding 2D-MOT chamber. 
It is attached to the science chamber via a differential pumping stage with a direct line of sight between the centers of the two MOT systems. 
The 2D-MOT is loaded with rubidium atoms from the background pressure and creates a transversally cooled beam of cold atoms guided to the MOT in the science chamber.

As cooling and trapping relies on (position and velocity dependant) radiation pressure due to light scattering, MOT operation requires a closed cooling cycle transition for the atoms. 
That is a state pair, which the atoms cannot leave during absorption or spontaneous emission. 
In $^{87}Rb$ such a transition is found in the D2 manifold between the $5S_{1/2}$ ground state and the $5P_{3/2}$ excited state, near \unit[780]{nm}.  
Both states are split in two respectively four hyperfine manifolds with total angular momentum $F=1,2$ and $F'=0,1,2,3$. 
We use the $5S_{1/2}, F=2 \rightarrow 5P_{3/2}, F'=3$ cooling transition, which is considered to be closed. 
For MOT operation, the cooling light is red detuned to this transition. 
Atoms can leave the otherwise closed cooling cycle only via off resonant excitation to the $5P_{3/2}, F'=2$ state and adjacent decay to the F=1 ground state. 
A repump laser resonant to the $5S_{1/2}, F=1 \rightarrow 5P_{3/2}, F'=2$ transition and overlapped with the cooling light is used to bring back the atoms to the cooling cycle. 
The repump beam is circularly polarised, has a power of \unit[20]{mW}, and a beam waist equal to the cooling beams. The 2D-MOT is operated with separate cooling and repump beams at the same frequencies.

\subsection*{Control Implementation}
The control of the experimental apparatus is divided between the real-time system J\"ager Adwin Pro II and a desktop PC dedicated for the RL agent.
While the Adwin system is responsible for the time-of-flight imaging operations, which require exact timing, the desktop PC controls the selected experimental parameters ($\Delta$ and $B'$) via a digital-to-analog converter (DAC) card (AdLink DAQ-2502).
The analog output line of the DAC card for the detuning $\Delta$ controls directly the offset lock used for frequency stabilization of the laser.
$B'$ is set via further analog output lines that control the current in the MOT-coils.

An episode is initiated from the PC via enabling the magnetic field gradient, effectively turning the MOT on.
An additional output line from the DAC card to the Adwin system is used to trigger the time-of-flight imaging sequence at the end of an episode.
At the conclusion of the imaging sequence, the Adwin system returns into the idle state, ready to start loading atoms. 

The fluorescence camera (UEye UI-3060CP) is connected to the PC and is constantly recording images and transferring them into a reserved memory register at a rate of \unit[150]{fps}.
During an episode, in each time step the latest available image is read out from the memory and processed to be passed as part of the observation to the agent.

\subsection*{Fluorescence imaging}
As atoms in the MOT scatter laser light during the cooling and trapping process, the scattered light can be observed as fluorescence.
This phenomenon makes the atomic cloud in the MOT observable, with the spatial distribution of atoms being reflected in the appearance of the fluorescence. 
Typically, the extension of the fluorescing cloud is on the order of millimeters. 
The rate at which fluorescence is emitted depends on the laser detuning, and is modulated by the magnetic field gradient.
This results in a highly non-linear mapping of the 3-dimensional atomic density distribution onto the imaging plane. 
The images then correspond to a projection along the imaging axis and show the fluorescence intensity in the direction of the camera. 
Different processes such as radiation trapping or imbalanced radiation pressure can lead to a variety of non-trivial shapes of the atomic cloud, including ring structures \cite{MOT_shapes93}. \\
The fluorescence images have a dimension of 50x50 pixels with 8-bit resolution. 
The exposure settings are chosen such that even weak signals can be detected, which is important for the initial phase of the learning. 
This leads to fast saturation when loading at maximum rate.

We have conducted tests in the simulated environment providing the agent only with integrated fluorescence signals instead of the full image as observable.
The training in such a scenario did not converge to the optimal policy, in particular, the agent was not able to adapt to any perturbation.

\subsection*{Reference episodes}
The reference episodes, in which the control sequence was manually set to the intuitive optimal policy, i.e. loading at optimal detuning and ramping the detuning to the maximal value in the last time step, allowed us to normalize the evolution of the performance (i.e. reward $R$), correcting for possible long term drifts of the experimental apparatus.
In practice it means that in each time step of a reference episode the actions of the agent were overwritten with predefined values while keeping the timing constraints of the agent, i.e. the \unit[60]{ms} step size.
The reference episodes were included during training (see below) and when evaluating a trained agent.

\begin{figure*}[ht!]
    \includegraphics[width=1\textwidth]{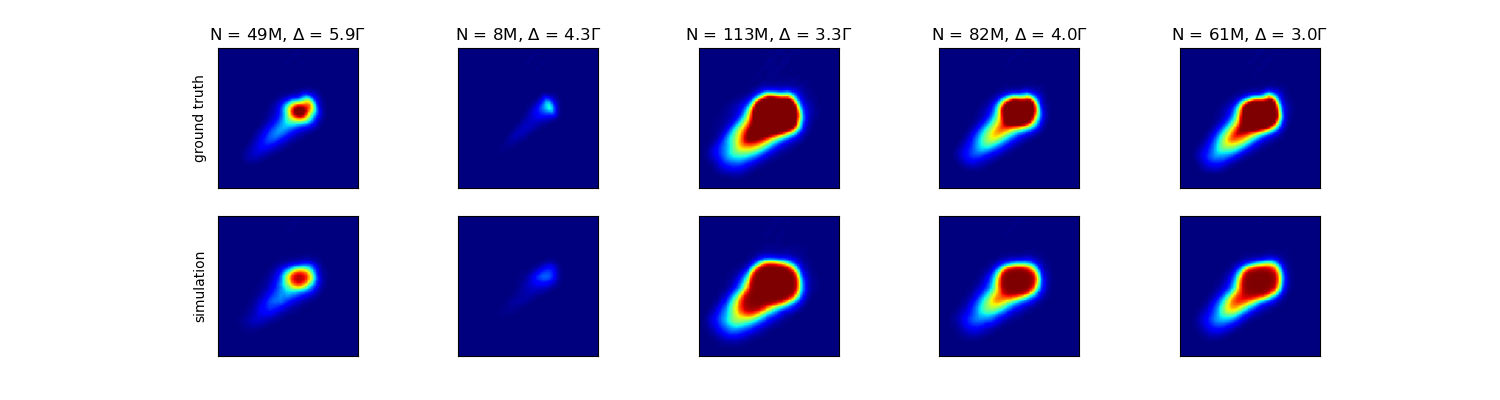}
	\caption{Experimental fluorescence images from the test dataset (upper line) and corresponding synthetic images (lower line) for different parameter sets ($N$, $\Delta$). }
\label{fig:generator_samples}
\end{figure*}

\subsection*{RL Algorithm}
For the realization of an RL-controlled MOT we have taken recourse to a modular framework known as \textit{coach} \cite{rlcoach}, containing implementations of a variety of RL algorithms in Python. 
In this work, we have focused on the DDPG algorithm, an actor-critic algorithm, that uses memory replay for training and is able to learn directly from raw images \cite{ddpg}.
We have found that almost all parameters are close to the optimum at their default values (as defined in \textit{coach}). 
The parameters used for the simulation can be found in our preset published as a part of the code \cite{RLMOT_code}.

Inspired by the performance of the Deep Q-Network \cite{Mnih2015} we have adapted the stacking of the 4 most recent fluorescence images as input to the agent.
This allows the agent to compute the rate of change of the fluorescence signal, which can be used to determine the loading rate of the MOT.
All values passed as observables are normalized to the range [0,1].

\subsection*{Scheduling}
The learning starts with a heat-up phase of 400 episodes during which the control is fully determined by a Brownian motion-like exploration. 
During the subsequent training phase, every 200 training episodes, a set of 10 reference and 10 evaluation episodes with random perturbation offsets are recorded. 
The performance of the agent is evaluated by disabling the exploration during the evaluation episodes.
During the training episodes, unlike in the original DDPG implementation, we compute a new action in every time step (i.e. no \textit{action repeat}). 

\subsection*{MOT simulation}
We have implemented a phenomenological, data-driven computer simulation for the MOT in order to better understand the behavior of the agent in such a novel environment, as well as to investigate sim-to-real transfer.

For our simulation we assume that the number of trapped atoms increases by a certain number $dN$ in each time step, where $dN$ only depends on the applied control parameter $\Delta_{n_t}$ at time step $n_t$.  
\begin{equation}
    N(n_t) = \sum_{i=1}^{n_t} dN\left( \Delta_i \right)\\
\end{equation}
This means  that we neglect loss processes that are known for MOTs and typically lead to saturation of the accumulated number of atoms.
Furthermore, we assume that at each time step the temperature of the trapped cloud is dictated by the cooling laser light and thus only depends on the value of the detuning. 
\begin{equation}
    T(n_t) = T\left( \Delta_{n_t} \right)
\end{equation}
This assumption is valid as long as the atomic cloud stays dilute and transparent to the light. 
Thus, we neglect density dependent multiple scattering.
In order to obtain the most accurate simulation, we experimentally determine the loading rate $dN$ as well as the temperature $T$ for a discrete number of detuning values and create look-up-tables (LUT), which are used for interpolations during the simulation (cf. Fig.~\ref{fig:4}\textbf{a}). 
The loading rate as function of the detuning is measured by setting the desired value $\Delta$ for a duration of {\unit[1]{s}} and measuring the final number of atoms via absorption imaging.
The temperature LUT is recorded by loading atoms at the optimal loading rate and subsequently rapidly ramping the detuning to a given value before performing absorption imaging.
The fluorescence images that are captured right before the measurement  of $dN$ and $T$  via absorption imaging are stored in a database. 
Each measurement is repeated 5 times. 

The full dataset contains on the order of 5000 samples.
The environment based on the simulation computes the intermediate values of $dN$  using a linear interpolation.
The temperature $T$ is obtained by interpolating between the measured values using an exponential function.
During the training on the simulation, noise is introduced to the loading rate $dN$ in order to simulate the physical fluctuation of the experimental conditions.
\begin{equation}
    dN \rightarrow dN \cdot\mathcal{N}\left(\mu,\sigma^2\right)
\end{equation}
with $\mathcal{N}$ being the normal distribution at mean value $\mu=1$ and standard deviation $\sigma=0.1$.

We trained a convolutional neural network (CNN) to generate realistically looking fluorescence images for the simulation on the same experimental dataset.
We have implemented the decoder-part of an autoencoder network \cite{autoencoder06,ConvAutoencoder11} as the generator and performed supervised training with $N$ and $\Delta$ as inputs and the fluorescence image as output.
The network consists of a fully connected layer, followed by 5 convolutional layers with rectified linear unit as activation function and upsampling layers in between.
Prior to the training we set fluorescence images to 0 for samples with $N=0$, as well as remove outliers from the dataset. 
A small, randomly chosen subset of the dataset was held out during training and used for testing the generator.
In Fig.~\ref{fig:generator_samples}, we compare the recorded images to the output of the generator for several parameter sets. 
The overall appearance of the generated images shows high similarity to the real images, with some fine grained details not fully captured by the CNN.

\subsection*{Data availability}
Data is fully available on request from the authors. Our code for training and evaluating an RL agent based on a simulated MOT can be found under \cite{RLMOT_code}.



\vspace{1em}
\noindent
\textbf{\large Acknowledgments}\\
The authors acknowledge fruitful discussions with G. Martius. 
We acknowledge funding from the German Research Foundation (DFG) through the Research Unit FOR 5413/1, Grant No. 465199066.

\vspace{1em}
\noindent
\textbf{\large Author contributions}\\
A.G., J.F., and V.V. planned and supervised the project. 
M.R. carried out the experiments.
V.V. carried out the simulations.
M.R., A.G. and V.V. analyzed the data.
M.R., A.G. and V.V. wrote the manuscript.
All authors discussed the results and contributed to the manuscript.

\vspace{1em}
\noindent
\textbf{\large Competing Interests}\\
The authors declare no competing interests.


\begin{thebibliography}{10}
\bibitem{note1}
\bibinfo{title}{Note, that the agent does not differentiate between training or
  evaluation episodes, since its actions solely depend on the obtained
  observations.}
\end{thebibliography}
\end{document}